# Generalized Uncertainty Principle in Quantum Gravity from Micro-Black Hole Gedanken Experiment

Fabio Scardigli[1]
Institute for Theoretical Physics,
University of Bern, Sidlerstrasse 5, CH-3012 Bern, Switzerland.

**Abstract**: We review versions of the Generalized Uncertainty Principle (GUP) obtained in string theory and in gedanken experiments carried out in quantum gravity. We show how a GUP can be derived from a measure gedanken experiment involving micro-black holes at the Planck scale of spacetime. The model uses only Heisenberg principle and Schwarzschild radius and is independent from particular versions of Quantum Gravity.

---

[1]postal address: via Europa 20, 20097 S. Donato Milanese, Milano, Italy.
e-mail: scardus@galactica.it

# 1. Generalized uncertainty principle in string theory

In string theory the Generalized Uncertainty Principle (GUP) has been well described since the works of G. Veneziano et al. [1] in 1986-90. There have been studied ultra high energy scatterings of strings in order to see how the theory tackles the inconsistences of quantum gravity at the Planck scale. The authors find interesting effects, new as regarding to those found in usual field theories, originating from the soft short-distance behaviour of string theory. They studied particularly the hard processes excitable at short distance as high-energy fixed-angle scatterings. They find that it is not possible to test distances shorter than the characteristic string length $\lambda_s = (\hbar \alpha)^{1/2}$ ($\alpha$ is the string tension). Another scale is dinamically generated, the gravitational radius (i.e. Schwarzschild) $R(E) \sim (G_N E)^{1/(D-3)}$ and the approaches towards $\lambda_s$ depend on whether or not $R(E) > \lambda_s$. If the latter is true, new contributions at distances of the order of R(E) appear, indicating a classical gravitational instability that can be attributed to black hole formation. If, on the contrary, $R(E) < \lambda_s$, those contributions are irrelevant: there are not black holes with a radius smaller than the string length. In this case, the analysis of short distances can go on and it has been shown that the larger momentum transfer do not always correspond to shorter distances. Precisely the analysis of the angle-distance relationship suggests the existence of a scattering angle $\theta_M$ such that when $\theta < \theta_M$ ($\theta$ = scattering angle) the relation between interaction distance and momentum transfer is the classical one (i.e. *a la* Heisenberg) with $<q> \sim \hbar / b$ (b=impact parameter), while, when $\theta >> \theta_M$ the classical picture is lost and becomes important a new regime where $<q> \sim b$. This suggests a modification of the uncertainty relation at the Planck scale in the form of

$$\Delta x \sim \frac{\hbar}{\Delta p} + Y\alpha \Delta p \qquad [1]$$

(where Y is a suitable constant) and consequently the existence of a minimal observable length of the order of string size $\lambda_s$. An analogous result has been obtained by Konishi et al. with an analysis based on the renormalization group. The proportionality constant Y depends on the particular version of string theory adopted.

The notion of minimal observable length has been described during the last years in the more modern language of duality [2]. An example is provided by the target-space duality (T-duality). It arises, for example, when the target space is a five dimensional manifold of the form $M_4 \times S^1$. It can be shown that the classical solution of the movement equations for the closed bosonic string compactified on the circle $S^1$ of radius R is

$$X^\mu(\tau,\sigma) = x_o^\mu + \alpha \frac{(\tau+\sigma)}{2}(\frac{n}{2R}+mR) + \alpha \frac{(\tau-\sigma)}{2}(\frac{n}{2R}-mR) +$$

$$+ i \sqrt{\frac{\alpha}{2}} \sum_{n \neq 0} ( \frac{a_n}{n} e^{in(\tau+\sigma)} + \frac{\overline{a_n}}{n} e^{-in(\tau-\sigma)} ) \qquad [2]$$

With the identification

$$p_L = \frac{\alpha}{2}(\frac{n}{2R}-mR); \quad p_R = \frac{\alpha}{2}(\frac{n}{2R}+mR) \quad [3]$$

this solution can be derived from a hamiltonian of type

$$H = \alpha^2(\frac{n^2}{4R^2}+m^2R^2) + oscillatory\ terms = 2(p_R^2+p_L^2) + oscill. \quad [4]$$

It can be noted that under the T duality

$$R \longleftrightarrow \frac{1}{2R}, \quad n \longleftrightarrow m \quad [5]$$

we have

$$p_R \rightarrow p_R; \quad p_L \rightarrow -p_L \quad [6]$$

The T duality can be interpreted as follows: a massless particle on a circle of radius R has a quantized momentum of p=n/2R. A string can also wrap m times around a circle with a momentum p=mR. The duality symmetry exchanges the two spectra, exchanging also R with $\alpha$/R. Simply, one cannot compress a circle below a certain length scale. The physical predictions of the theory are invariant under the replacement of the radius R of the fifth dimension with $\alpha$/2R. Thus, we cannot differentiate physically between a very small and a very large radius for the additional dimension. This invariance suggests the existence of a minimum length $R_{min} \sim (\alpha)^{1/2}$, and can be generalized to more than one extra dimension with a topology that is more complex than just a product of circles. Strings do not see spacetime in the same way as do point particles. When one probes small distances on the string scale, instead of probing them one just watches the propagation of larger strings.
On these grounds we should say that in string theory the very notion of ordinary continuum spacetime ceases to make sense below the minimum length of the string scale $\lambda_s$.
This picture has been quite deeply modified by the work of Shenker et al. [3]. It seems to be possible to probe spacetime also below the string scale. In fact previous results in string theory, which found evidence that the minimum length is the string scale, were based on the use of strings themselves as probes. The new approach uses D branes as probes and it should be able to test the behavior of string theory at distances far shorter than the string scale $\lambda_s$.

## 2. GUP in black hole gedanken experiments

Several other types of analysis have been performed during the last years about uncertainty relations and measurability bounds in quantum gravity (see e.g. [4], [5]). All of them are characterized by a 'deformation' of the classical Heisenberg uncertainty relation. Among the more interesting works is that of M.Maggiore [6]. He obtains an expression of a GUP by analyzing a gedanken experiment for the measurement of the area of the apparent horizon of a black hole in quantum gravity. This rather model-independent approach provides a GUP which agrees in the functional form with the similar result obtained in the framework of string theory.

The gedanken experiment proceeds by observing the photons scattered by the studied black hole. The main physical hypothesis of the experiment is that the black hole emits Hawking radiation. Recording many photons of the Hawking radiation we are able to obtain an 'image' of the black hole. Besides, measuring the direction of the propagation of photons emitted at different angles and tracing them back, we can (in principle) locate the position of the center of the hole. In this way we make a measure of the radius $R_h$ of the horizon of the hole. This measure suffers two kinds of errors. The first one is, as in Heisenberg classical analysis, the resolving power of the microscope

$$\Delta x^{(1)} \sim \frac{\lambda}{\sin\theta} \qquad [7]$$

where $\theta$ is the scattering angle. Besides during the emission process the mass of the black hole varies from $M+\Delta M$ to $M$ (with $\Delta M = h/c\lambda$) and the radius of the horizon changes accordingly. The corresponding error is intrinsic to the measurement and it values

$$\Delta x^{(2)} \sim \frac{2G}{c^2} \Delta M \qquad [8]$$

or

$$\Delta x^{(2)} \sim \frac{2G}{c^3} \frac{h}{\lambda} \qquad [9]$$

By means of the obvious inequality $\frac{\lambda}{\sin\theta} \geq \lambda$, the errors $\Delta x^{(1)}$ and $\Delta x^{(2)}$ are combined linearly (this step contains some arbitrariness) to obtain

$$\Delta x \geq \lambda + k \frac{2G}{c^3} \frac{h}{\lambda} \sim \frac{h}{\Delta p} + k \frac{2G}{c^3} \Delta p \qquad [10]$$

The numerical constant k cannot be predicted by the model-independent arguments presented. We shall develop in the present work an approach based on a gedanken experiment involving a planckian micro-black hole. This approach will allow us to obtain a GUP without the use of Hawking effect or other refined effects: we use only Heisenberg relation and the notion of gravitational (i.e. Schwarzschild) radius.

The notion of GUP has been further investigated by several authors by linking it to the deformed Poincare' algebra. The mathematical structure that encodes the GUP in a natural way is that of a deformed Heisenberg algebra

$$[X_i, X_j] = -\frac{\hbar^2}{4\kappa^2} i\, \epsilon_{ijk} J_k$$

$$[X_i, P_j] = i\hbar\, \delta_{ij} \left(1 + \frac{P^2 + m^2}{4\kappa^2}\right)^{1/2}$$

[11]

This gives an interpretation of the GUP at a purely kinematical level, independently from any specific dynamical theory.
From the preceding equation the GUP follows

$$\Delta x_i \Delta p_j \geq \frac{\hbar}{2} \delta_{ij} \left\langle \left(1 + \frac{P^2 + m^2}{4\kappa^2}\right)^{1/2} \right\rangle \sim \frac{\hbar}{2} \delta_{ij} \left(1 + \frac{(\Delta p)^2}{8\kappa^2} + \frac{p^2 + m^2}{8\kappa^2}\right)$$

[12]

where we have used $\langle \mathbf{P}^2 \rangle = \mathbf{p}^2 + (\Delta p)^2$ and we suppose to be in the regime $p^2 + m^2 \ll \kappa^2$ and $\Delta p \leq \kappa$.

## 3. GUP from a micro-black hole gedanken experiment

In the present section we want to derive an expression of a GUP from the analysis of the measure process carried out in presence of gravity. The basic idea is that at Planck level spacetime admits great fluctuations of the metric and therefore the possibility of many (virtual) micro-black holes. This structure has been referred to as a spacetime foam structure and it has been discussed in innumerable papers (see e.g. [7]) and employed to calculate many different properties in the quantum theory of gravity. For example it serves as intuitive physical model for the calculus of the microscopic origin of black hole entropy (see e.g. [8]).
We shall apply the Heisenberg principle to the measure process and we shall show how the formation of a micro-black hole affects the measure process itself. Starting from the principle written in the form $\Delta p \Delta x \geq \hbar / 2$ and reminding that in our high energy situation $\Delta E \sim c \Delta p$, we can cast the Heisenberg inequality[2] in the form $\Delta E \Delta x \geq \hbar c / 2$.

When we observe a space region of width $\Delta x$, we should expect that the metric field in that region undergoes quantum fluctuations with an amplitude in energy of $\Delta E \sim \hbar c / 2\Delta x$. This energy is confined in the region of width $\Delta x$.

---

[2] We remind of course that only p and x are conjugate operators in Hilbert space and that the relation $\Delta E \Delta x \geq \hbar c/2$ must be seen as a simple algebraic tool without any operatorial meaning.

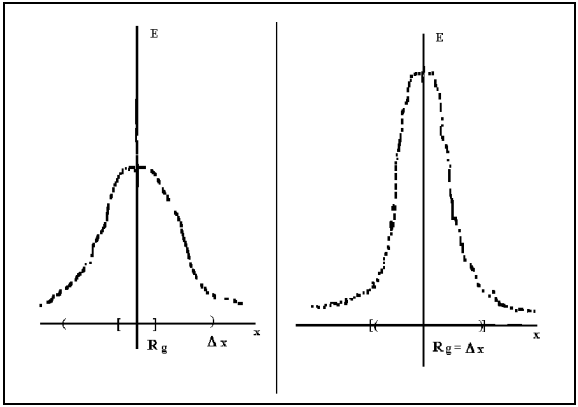

The gravitational radius $R_g = \dfrac{2G\Delta E}{c^4}$ associated with the energy $\Delta E$ falls usually well inside the region $\Delta x$. But when we shrink the observed region, the fluctuation $\Delta E$ grows up and the corresponding $R_g$ becomes wider, untill it reaches the same width of $\Delta x$ (see fig.1). It is easy to show that this critical length is the Planck length [9] and the associated energy is the Planck energy $\varepsilon_P$. A micro-black hole originates. If one want to observe more refined details, he should concentrate in that region an energy greater than the Planck energy $\varepsilon_P$ and this would enlarge further the gravitational radius $R_g$, hiding in this way more details of the region beyond the event horizon of the microhole.

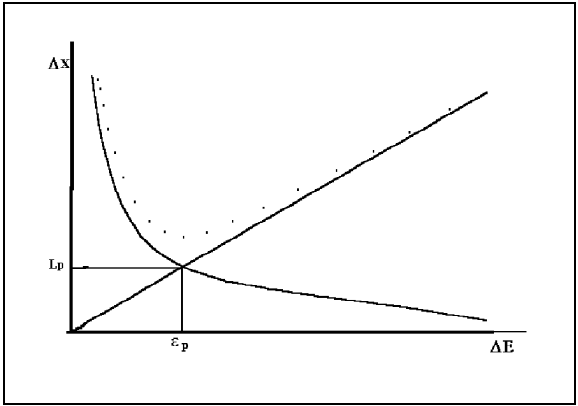

It emerges clearly that the minimum observable region has a width equal to $L_P$. Every more refined detail is hidden by the event horizon and every effort to shrink further the region $\Delta x$ would result in a greater quantum fluctuation of the energy and therefore of the amplitude of the unobservable region. Because $R_g$ goes linearly with $\Delta E$, while $\Delta x$ goes with an inverse proportion with $\Delta E$, we could summarize the situation in the diagram of fig. 2.

We have

$$\Delta x \geq \begin{cases} \dfrac{\hbar c}{2\Delta E} & \text{for} \quad \Delta E \leq \epsilon_P \\[2ex] \dfrac{2G\Delta E}{c^4} & \text{for} \quad \Delta E > \epsilon_P \end{cases} \qquad [13]$$

The simplest way to combine the preceding inequalities into a single one is the linear way

$$\Delta x \geq \frac{\hbar c}{2\Delta E} + \frac{2G\Delta E}{c^4} \qquad [14]$$

that, in term of $\Delta p$, reads

$$\Delta x \geq \frac{\hbar}{2\Delta p} + \frac{2G\hbar}{c^3}\frac{\Delta p}{\hbar} = \frac{\hbar}{2\Delta p} + 2L_p^2\frac{\Delta p}{\hbar} \qquad [15]$$

We obtain in this way a generalization of the uncertainty principle to cases in which the gravity is important and to energy of the order of $\varepsilon_P$.

We should say that in the present gedanken experiment the continuity of spacetime holds up to any length. Spacetime is continuous near or inside the microhole and when the microhole has evaporated spacetime behaves yet as a continuum. The GUP can be obtained without renouncing to the continuity. There is a difference between this approach and several other visions which present spacetime as a lattice: e.g. naive spacetime foam; certain interpretations of spacetime, given by loop quantum gravity, as a polymeric structure; quantum geometry (topological transitions). The continuity of spacetime at every scale is in agreement with the recent results [3] obtained in string theory.

This derivation is based on the Heisenberg principle (which is supposed to be valid untill the Planck length is reached) and on the Schwarzschild radius, that again derives from the finiteness of the speed of light. This derivation is independent from particular versions of quantum gravity or from gravitational effects in curved spacetime (e.g. Hawking effect) and also from particular models of the micro-objects (strings etc.).

## 4. Discussion and conclusion

Several interesting notes can be made about the GUP obtained in the previous section and the micro-black hole mechanism examinated. One of the most intriguing is the following. It is well known that we cannot look inside the region occupied by a black hole, because of the presence of the event horizon, and this situation persists for all the lifetime of the hole. Precisely, if there were not any gravitational effect and the Heisenberg classical uncertainty principle were valid also beyond the Planck scale , we would expect that the Planck energy $\varepsilon_P$ , concentrated in a region of amplitude $L_P$ , take a time equal to $\frac{L_P}{c} = \tau_P = \frac{\hbar/2}{\epsilon_P}$ to escape from the region considered. That region would be re-testable after a time $\tau_P$ (for example, with a light flash). However, the presence of gravity during the measure process causes the formation of a microhole and therefore of an horizon. The horizon does not permit to the energy to escape from the hole (apart from the little amount of energy escaping away by Hawking evaporation). The region inside the microhole remains therefore unobservable for a time very much longer than the time the light takes to travel a Planck distance. The time during which the region inside a microhole remains unobservable (or untestable) is clearly equal to the lifetime of the planckian micro hole itself: a strightforward calculation, based on the Hawking evaporation, gives for it a time of about 2000 $\tau_P$ ($\tau_P$ = Planck time). There is a sort of 'plasticity' of spacetime, which seems to be able to 'remember' for a long time the 'too large' deformations suffered (that is those causing the formation of an event horizon). We can summarize the situation by introducing an "UnObservability Principle", settling that a region of width L_p cannot be observed, or tested (after a first observation), not even in principle, for a time equal to

$$\tau_{TOT} = \tau_{Planck} + \text{\textit{lifetime of a planckian black hole.}} \qquad [16]$$

This un-observability time results to be longer than that we would have if the gravity did not exist. Of course this considerations are true if the semiclassical formulae (i.e. Hawking radiation) here employed keep their validity also at the Planck level.
Finally we note that a microhole lifetime of the order of $\sim 10^3 \; \tau_P$ implys a wideness in energy of the order of $\Delta E \sim (\hbar / 4000 \; \tau_P) \sim 10^{16}$ GeV. This result is very interesting and it is in contrast with the more common views. It seems to indicate that there could be micro holes also at energies less than the Planck threshold. $10^{16}$ GeV is quite similar to the Grand Unification Energy. So quantum-gravitational effects may be far closer in energy to every day physics than previously believed. Quite astonishingly, a similar result has been obtained, in a totally different way, by calculations in 11-dimensional M-theory [10]. May be quantum gravity could be nearer than we have ever thought !

## 5. Acknowledgements

The author wishes to thank Juan Maldacena and the Anonymous Referee to have drawn his attention on the paper in reference [3].